\documentclass[aps,pra,twocolumn,showpacs,groupedaddress,preprintnumbers,amsmath,amssymb]{revtex4-1}
\usepackage{mathrsfs}

\usepackage{graphicx}
\usepackage{bm}
\usepackage{float}

\begin{document}

\preprint{APS/123-QED}

\title{Scalable one-way quantum computer using on-chip resonator qubits}

\author{Chun-Wang Wu}
  \email{cwwu@nudt.edu.cn}
\author{Ming Gao}
\author{Hong-Yi Li}
\author{Zhi-Jiao Deng}
\author{Hong-Yi Dai}
\author{Ping-Xing Chen}
\author{Cheng-Zu Li}
\affiliation{College of Science, National University of Defense
 Technology, Changsha 410073, People's Republic of China}

\date{\today}

\begin{abstract}
 We propose a scalable and robust architecture for one-way quantum computation using coupled networks of superconducting
 transmission line resonators. In our protocol, quantum information is encoded into the long-lived photon states of
 the resonators, which have a much longer coherence time than the usual superconducting qubits. Each resonator contains
 a charge qubit used for the state initialization and local projective measurement of the photonic qubit. Any pair of
 neighboring photonic qubits are coupled via a mediator charge qubit, and large photonic cluster states can be created by
 applying Stark-shifted Rabi pulses to these mediator qubits. The distinct advantage of our architecture is that it combines
 both the excellent scalability of the solid-state systems and the long coherence time of the photonic qubits. Furthermore,
 this architecture is very robust against the parameter variations.
\end{abstract}


\maketitle

\section{introduction}

Entanglement lies at the heart of quantum information processing \cite{reve1}. In 2001, Raussendorf and Briegel showed that
a special type of highly entangled multiqubit states, called cluster states, can be used to implement one-way quantum
computation \cite{reve2}. In contrast to the standard quantum circuit model, which uses single- and two-qubit logic gates,
a one-way quantum computer proceeds by a sequence of single-qubit measurements with classical feedforward of their outcomes.
Moreover, this new kind of quantum computation is universal in the sense that any quantum circuit can be implemented on a
suitable cluster state by single-qubit measurements only.

Quantum computation on cluster states has been studied in a variety of physical systems. To date, a small-scale one-way quantum
computation has been demonstrated using linear optics techniques \cite{reve3,reve4,reve5}. As quantum information carriers,
photonic qubits have the advantage of long coherence time. However, it is hard to construct a scalable optical one-way quantum
computer due to the difficulty of large-scale integration in the linear optical devices. One-way quantum computation has also been explored
in artificial solid-state systems (e.\,g.\,electron spins in quantum dots \cite{reve6,reve7} and superconducting qubits \cite
{reve8,reve9,reve10}). Thanks to the well-established microfabrication techniques, these solid-state qubits have very excellent
scalability, but their inherent bad coherence properties remain a stumbling block. A physical architecture, which combines both
the scalability and long coherence time, is desirable for the realization of scalable one-way quantum computation.

Here, we propose an alternative architecture for one-way quantum computation using on-chip resonator qubits, which to some extent
can overcome the two major roadblocks -- decoherence and scalability -- at the same time. In our protocol, quantum information is encoded
into the zero- and one-photon states of the high-Q transmission line resonators. The good coherence properties of these photon states
have been demonstrated in recent experiments. Many interesting complex quantum states of photons, including the arbitrary superposition
of Fock states \cite{reve11} and the high NOON states \cite{reve12}, have been synthesized in the laboratory. More recently, researchers
also proposed using these photon states to implement quantum computation based on the standard circuit model \cite{reve13,reve14,reve15}.
Each resonator in our architecture contains a charge qubit inside it, which is used for the state initialization and local projective
measurement of the photonic qubit. The controlled-phase interaction between neighboring photonic qubits can be implemented by applying
$2\pi$ Stark-shifted Rabi pulses to the charge qubits at the resonator junctions, and large photonic cluster states can be created in $2d$ steps
where $d$ is the dimension of the resonator lattice. Moreover, our architecture is very robust against the parameter variations.

\begin{figure*}[!t]
 \includegraphics[scale=0.75]{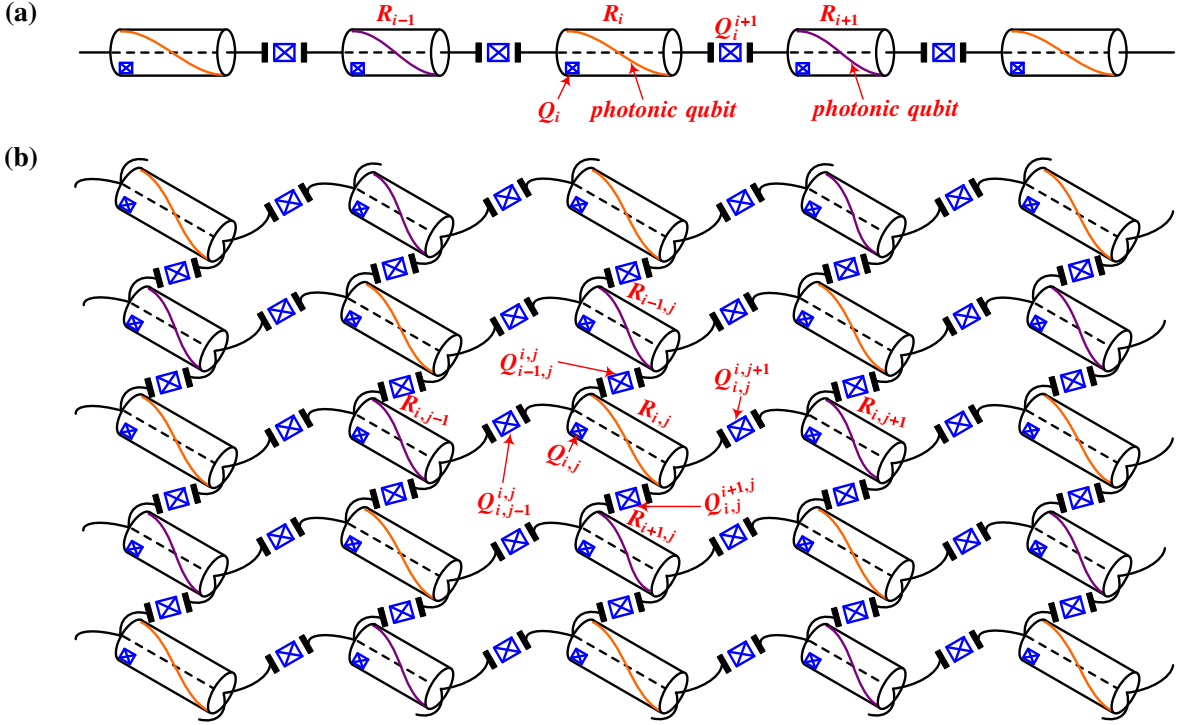}
 \setlength{\belowcaptionskip}{-0.4cm}
 \caption{\label{fig1}(Color online) Schematic layout of our proposed architecture for one-way quantum computation based on the 1D linear resonator
 array (a) and 2D resonator square lattice (b). Quantum information is encoded into the zero- and one-photon states of the high-Q microwave resonators.
 Each resonator contains a charge qubit used for the state initialization and local projective measurement of the photonic qubit. Any pair of neighboring resonators
 have different resonance frequencies and are simultaneously capacitively coupled to a mediator charge qubit. }
 \end{figure*}

The paper is organized as follows. In Sec.\ \uppercase\expandafter{\romannumeral2} we describe our proposed architecture for one-way quantum
computation in detail. In Sec.\ \uppercase\expandafter{\romannumeral3}, a concrete experimental procedure is presented including how to initialize
the system, create the large photonic cluster states, and implement the local projective measurements of photonic qubits in the arbitrary basis.
Finally, the experimental feasibilities are analyzed in Sec.\ \uppercase\expandafter{\romannumeral4}.

 \section{physical architecture}

 In principle, our idea is applicable for implementing one-way quantum computation based on an arbitrary dimensional resonator array. In Fig.\,1,
 we only draw the one-dimensional (1D) linear resonator array and two-dimensional (2D) resonator square lattice as simple examples.

 For the 1D case, a linear array of $n$ transmission line resonators $R_{1},\;R_{2},\;...\;,\;R_{n}$ are connected via $n-1$ superconducting charge qubits
 $Q^{2}_{1},\;Q^{3}_{2},\; ...\;,\;Q^{n}_{n-1}$ (Fig.\,1(a)). Any pair of neighboring resonators $R_{i}$ and $R_{i+1}$ ($1\leq i\leq n-1$) are required to have
 different resonance frequencies. Here, we assume that all the resonators $R_{p}$ ($1\leq p\leq n$, $p$ is odd) have the same resonance frequency $w$,
 while the other resonators $R_{q}$ ($1\leq q\leq n$, $q$ is even) have the resonance frequency $w^{'}$ ($w^{'}\neq w$). Then the Hamiltonian that describes
 the photons in the resonator modes is (assuming $\hbar=1$)
 \begin{equation}
 H_{1}^{1D}=\sum_{1\leq p\leq n}^{p\; is\; odd}wa^{\dag}_{p}a_{p}+\sum_{1\leq q\leq n}^{q\; is\; even}w^{'}a^{\dag}_{q}a_{q},
 \end{equation}
 where $a^{\dag}_{p}$ ($a^{\dag}_{q}$) and $a_{p}$ ($a_{q}$) are the photon creation and annihilation operators for resonator $R_{p}$ ($R_{q}$). In our scheme,
 the photon number of any resonator at any time is engineered to be smaller than 2, and we use the zero-photon state $|0\rangle_{i}$ and one-photon state $|1\rangle_{i}$
 of resonator $R_{i}$ to represent the two states of logical qubit $i$. This kind of photonic qubit has a much longer coherence time than the usual superconducting
 qubits. Each resonator $R_{i}$ contains a charge qubit $Q_{i}$ used for the single-qubit rotations and fast readout of the photonic qubit $i$. $Q_{i}$'s role will be
 discussed detailedly later. The neighboring resonators $R_{i}$ and $R_{i+1}$ are simultaneously capacitively coupled to a mediator charge qubit $Q_{i}^{i+1}$. Let us denote the lowest two eigenstates of qubit $Q_{i}^{i+1}$ with $|g\rangle_{i}^{i+1}$ and $|e\rangle_{i}^{i+1}$, which are separated by energy
 $\epsilon_{i}^{i+1}$ and coupled to its adjacent resonators with qubit-resonator coupling strength $g$. Then the interaction between the resonators and the mediator qubits at the resonator junctions can be described by
 \begin{eqnarray}
 H_{2}^{1D}&=&\sum_{i=1}^{n-1}[\epsilon_{i}^{i+1}|e\rangle_{i}^{i+1}\langle e|_{i}^{i+1}+g(a_{i}^{\dag}|g\rangle_{i}^{i+1}\langle e|_{i}^{i+1} \nonumber \\
 &&+a_{i+1}^{\dag}|g\rangle_{i}^{i+1}\langle e|_{i}^{i+1}+H.c.)].
 \end{eqnarray}
 We assume that control of $Q_{i}^{i+1}$ can be exercised by a ``shift'' pulse to tune $\epsilon_{i}^{i+1}$ or a resonant microwave pulse to induce Rabi oscillation between $|g\rangle_{i}^{i+1}$ and $|e\rangle_{i}^{i+1}$.

 For the 2D case, $n^{2}$ resonators $R_{i,j}$ ($1\leq i,j\leq n$) are connected by $2n(n-1)$ charge qubits $Q_{i,j}^{i^{'},j^{'}}$ ($1\leq i,j,i^{'},j^{'}\leq n$,
 $i^{'}+j^{'}-i-j=1$) to form an $n\times n$ square lattice (Fig.\,1(b)). To suppress the photon hopping between neighboring resonators, we arrange that all the resonators $R_{p,q}$ ($1\leq p,q\leq n$, $p+q$ is even) have the same resonance frequency $w$, while the other resonators $R_{k,m}$ ($1\leq k,m\leq n$, $k+m$ is odd) have the resonance frequency $w^{'}$. Then the Hamiltonian of the resonator modes reads
 \begin{equation}
 H_{1}^{2D}=\sum_{1\leq p,q\leq n}^{p+q\; is\; even}wa^{\dag}_{p,q}a_{p,q}+\sum_{1\leq k,m\leq n}^{k+m\; is\; odd}w^{'}a^{\dag}_{k,m}a_{k,m},
 \end{equation}
 where $a^{\dag}_{p,q}$ ($a^{\dag}_{k,m}$) and $a_{p,q}$ ($a_{k,m}$) are the photon creation and annihilation operators for resonator $R_{p,q}$ ($R_{k,m}$). Each resonator $R_{i,j}$ contains a charge qubit $Q_{i,j}$. The mediator qubit $Q_{i,j}^{i^{'},j^{'}}$'s lowest two eigenstates $|g\rangle_{i,j}^{i^{'},j^{'}}$ and $|e\rangle_{i,j}^{i^{'},j^{'}}$, separated by energy $\epsilon_{i,j}^{i^{'},j^{'}}$, are simultaneously capacitively coupled to $R_{i,j}$ and $R_{i^{'},j^{'}}$ with coupling strength $g$. Then the Hamiltonian that describes the interaction between the resonators and the mediator qubits is
 \begin{eqnarray}
 &&H_{2}^{2D}=\sum_{1\leq i,j\leq n}[\epsilon_{i,j}^{i,j+1}|e\rangle_{i,j}^{i,j+1}\langle e|_{i,j}^{i,j+1}+\epsilon_{i,j}^{i+1,j}|e\rangle_{i,j}^{i+1,j}\langle e|_{i,j}^{i+1,j} \nonumber \\
 &&+g(a^{\dag}_{i,j}|g\rangle_{i,j}^{i,j+1}\langle e|_{i,j}^{i,j+1}+a_{i,j+1}^{\dag}|g\rangle_{i,j}^{i,j+1}\langle e|_{i,j}^{i,j+1}+H.c.)  \nonumber \\
 &&+g(a^{\dag}_{i,j}|g\rangle_{i,j}^{i+1,j}\langle e|_{i,j}^{i+1,j}+a_{i+1,j}^{\dag}|g\rangle_{i,j}^{i+1,j}\langle e|_{i,j}^{i+1,j}+H.c.)],
 \end{eqnarray}
 where $\epsilon_{i,n}^{i,n+1}=\epsilon_{n,j}^{n+1,j}=0$ ($1\leq i,j \leq n$), and $a_{i,n+1}=a_{n+1,j}=0$ ($1\leq i,j \leq n$).

 In principle, the structures drawn in Fig.\,1 can be extended to the higher dimensional case. In Ref.\,\cite{reve16}, the authors showed that, it is possible to realize
 an effective arbitrary dimensional resonator lattice system by appropriately engineering the connections between the resonators fabricated on a 2D chip. However, with the increasing dimension, we must deal with the experimental problem of overlapping connections. Maybe it can be solved by fabricating crossing lines in different
 layers \cite{reve17}. In fact, a 2D resonator square lattice suffices to implement the universal one-way quantum computation and no overlapping connections are needed
 for this case.

 In our architecture, the coupling between any pair of adjacent resonators is independently tunable. By tuning the resonance frequency of a mediator charge qubit and making it far detuned from its neighboring resonators, the interaction between the two resonator qubits can be switched off. This is essential for the implementation
 of one-way quantum computation. First, once the photonic cluster states are generated, the interqubit coupling should be disabled to prevent the system state from further evolution. Second, an isolated photonic qubit is convenient for preparing the initial state and performing the local projective measurements.

 Large cluster states can be generated by ``fusing'' the neighboring qubits via conditional phase gates \cite{reve2}. In our architecture, the controlled-phase interaction between a pair of neighboring photonic qubits can be implemented by applying a $2\pi$ Stark-shifted Rabi pulse to the mediator charge qubit (Fig.\,2). Now, we will explain this method based on the linear resonator array. We assume that the two involving resonators, $R_{i}$ with resonance frequency $w$ and $R_{i+1}$ with resonance frequency $w^{'}$, are isolated from the other resonators (by tuning the transition frequencies of $Q_{i-1}^{i}$ and $Q_{i+1}^{i+2}$ instantaneously to the far detuned
 regime), and their mediator charge qubit, $Q_{i}^{i+1}$, is operated in the strong dispersive regime. In this case, the qubit transition of $Q_{i}^{i+1}$ can be resolved into separate spectral lines for different photon number states of $R_{i}$ and $R_{i+1}$. Corresponding to $n$ photons in $R_{i}$ and $n^{'}$ photons in $R_{i+1}$, the Stark-shifted transition frequency of $Q_{i}^{i+1}$ is \cite{reve18}
 \begin{equation}
 \epsilon_{i}^{i+1}(n;n^{'})=\epsilon_{i}^{i+1}+\frac{g^{2}}{\epsilon_{i}^{i+1}-w}(2n+1)+\frac{g^{2}}{\epsilon_{i}^{i+1}-w^{'}}(2n^{'}+1).
 \end{equation}
 An additional microwave field with frequency $w_{d}$ is applied to $Q_{i}^{i+1}$, which is described by
 \begin{equation}
 H_{drive}=\Omega(|e\rangle_{i}^{i+1}\langle g|_{i}^{i+1}e^{-iw_{d}t}+H.c.),
 \end{equation}
 where $\Omega$ is the Rabi strength. Now we choose $w_{d}=\epsilon_{i}^{i+1}(1;1)$ and $|\Omega|\ll\frac{2g^{2}}{|\epsilon_{i}^{i+1}-w|}$, $\frac{2g^{2}}{|\epsilon_{i}^{i+1}-w^{'}|}$, then the mediator charge qubit $Q_{i}^{i+1}$ will undergo Rabi oscillations if both $R_{i}$ and $R_{i+1}$ have one photon in them, but do nothing for other photon states. With the choice of $\Omega t=\pi$, the system state evolution is
 \begin{equation}
 \left\{
 \begin{array}{ccc}
 |0\rangle_{i}|0\rangle_{i+1}|g\rangle_{i}^{i+1}&\rightarrow&|0\rangle_{i}|0\rangle_{i+1}|g\rangle_{i}^{i+1},\\
 |1\rangle_{i}|0\rangle_{i+1}|g\rangle_{i}^{i+1}&\rightarrow&|1\rangle_{i}|0\rangle_{i+1}|g\rangle_{i}^{i+1},\\
 |0\rangle_{i}|1\rangle_{i+1}|g\rangle_{i}^{i+1}&\rightarrow&|0\rangle_{i}|1\rangle_{i+1}|g\rangle_{i}^{i+1},\\
 |1\rangle_{i}|1\rangle_{i+1}|g\rangle_{i}^{i+1}&\rightarrow&-|1\rangle_{i}|1\rangle_{i+1}|g\rangle_{i}^{i+1}.\\
 \end{array} \right.
 \end{equation}
 By tracing out the auxiliary system $Q_{i}^{i+1}$, we actually obtain a conditional phase gate between photonic qubits $i$ and $i+1$. The Stark-shifted Rabi oscillation used in this method has been experimentally demonstrated \cite{reve19,reve20}, and more recently was used in an entangled state synthesis algorithm \cite{reve21}.
 \begin{figure}[!t]
 \includegraphics[scale=0.65]{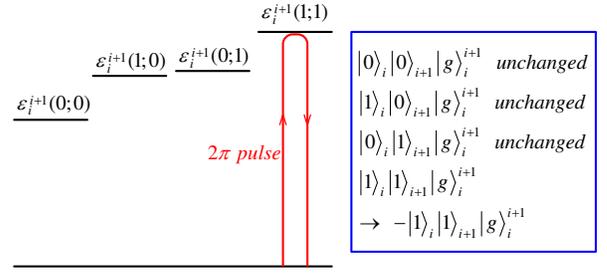}
 \setlength{\belowcaptionskip}{-0.4cm}
 \caption{\label{fig2}(Color online) The conditional phase gate between a pair of neighboring photonic qubits can be implemented by applying a $2\pi$ Stark-shifted Rabi
 pulse to the mediator charge qubit.}
 \end{figure}

 Local operations of a photonic qubit can be performed with the help of a charge qubit injected into the resonator. High-Q resonator modes are advantageous for the quantum information encoding but are adverse to the local measurements. To solve this problem, we can use the technique of engineering two modes of a resonator with different quality factors, which has been demonstrated experimentally in \cite{reve22}. Now, we introduce the detailed configuration inside $R_{i}$. As shown in Fig.\,3, the charge qubit $Q_{i}$, fabricated at one end of $R_{i}$, is capacitively coupled to the resonator modes. The transition between $Q_{i}$'s lowest two eigenstates $|g\rangle_{i}$ and $|e\rangle_{i}$, which are separated by energy $\epsilon_{i}$, can be driven by applying a microwave pulse $U_{i}^{d}$ to the gate. $\epsilon_{i}$ can be controlled using a local flux bias line by changing the applied magnetic flux $\Phi_{i}$. The high-Q half-wave mode of $R_{i}$ is used for encoding the quantum information, and $R_{i}$'s low-Q full-wave mode is strongly coupled to a measurement line fabricated at the resonator center. The state of $Q_{i}$ can be measured by tuning $\epsilon_{i}$ close to the full-wave resonance frequency of $R_{i}$ and applying a microwave field $U_{i}^{m}$ at the input port of the measurement line \cite{reve22}.
 \begin{figure}[!t]
 \includegraphics[scale=0.64]{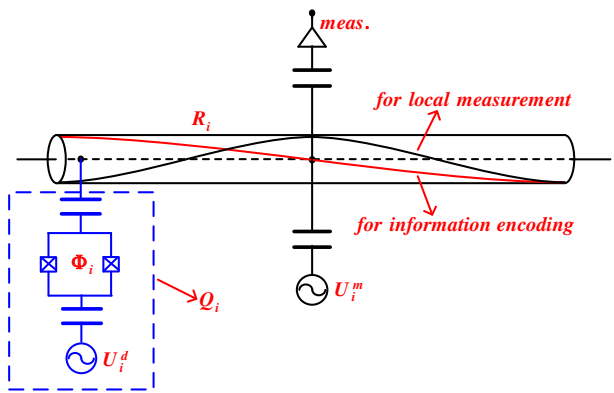}
 \setlength{\belowcaptionskip}{-0.4cm}
 \caption{\label{fig3}(Color online) Detailed configuration inside $R_{i}$. The charge qubit $Q_{i}$ is fabricated at one end of $R_{i}$ and capacitively coupled to the resonator modes. The high-Q half-wave mode of $R_{i}$ is used for the quantum information encoding, while $R_{i}$'s low-Q full-wave mode is strongly coupled to a measurement line fabricated at the resonator center. }
 \end{figure}

 \section{one-way quantum computation using on-chip resonator qubits}

 In this section, we give a concrete experimental procedure to implement one-way quantum computation based on our proposed architecture. We will describe in detail the
 total manipulation process including state initialization, creation of the large photonic cluster states, and implementation of the local projective measurements in the
 arbitrary basis.

 To generate the large photonic cluster states, the photonic qubits should be prepared in the state $\otimes_{i}\frac{1}{\sqrt{2}}(|0\rangle_{i}+|1\rangle_{i})$. Initially, we assume the composite system of $Q_{i}$ and $R_{i}$ is in the ground state $|g\rangle_{i}|0\rangle_{i}$, and $Q_{i}$ is detuned from $R_{i}$. Then, by
 applying a $\frac{\pi}{2}$-pulse to $Q_{i}$ through the local driving line, the system state can be driven into $\frac{1}{\sqrt{2}}(|g\rangle_{i}-i|e\rangle_{i})|0\rangle_{i}$. Next, we tune the transition frequency of $Q_{i}$ to be resonant with $R_{i}$'s half-wave mode instantaneously. After a time duration $t=\frac{3\pi}{2g_{i}}$, where $g_{i}$ is the qubit-resonator coupling strength between $Q_{i}$ and $R_{i}$, the system state evolves into $\frac{1}{\sqrt{2}}|g\rangle_{i}(|0\rangle_{i}+|1\rangle_{i})$, and $Q_{i}$ is tuned to be detuned from $R_{i}$ again. Since each resonator has its own auxiliary charge qubit and driving line, parallel operation is allowed for this state initialization process.

 For the 1D case, we show how to create the large photonic cluster state from the initial state $\otimes_{i=1}^{n}\frac{1}{\sqrt{2}}(|0\rangle_{i}+|1\rangle_{i})$. In the first step, the photonic qubit pairs 1 and 2, 3 and 4, 5 and 6, $\ldots$ are isolated to form independent subsystems by tuning the transition frequencies of $Q_{2}^{3}$, $Q_{4}^{5}$, $Q_{6}^{7}$, $\ldots$ to the far detuned regime instantaneously, then we ``fuse'' the qubits 1 and 2, 3 and 4, 5 and 6, $\ldots$ using conditional phase gates by applying $2\pi$ Stark-shifted Rabi pulses to $Q_{1}^{2}$, $Q_{3}^{4}$, $Q_{5}^{6}$, $\ldots$ . After this step, the state of all the photonic qubits is
 \begin{equation}
 \frac{1}{2^{\frac{n}{2}}}\otimes_{i=1,3,5,\ldots}(|0\rangle_{i}+\sigma_{i+1}^{z}|1\rangle_{i})(|0\rangle_{i+1}+|1\rangle_{i+1}),
 \end{equation}
 where $\sigma_{i+1}^{z}$ is the Pauli-Z operator for qubit $i+1$. In the second step, we isolate photonic qubit pairs 2 and 3, 4 and 5, 6 and 7, $\ldots$ and perform conditional phase gates for each pair as in the first step. Then we can prepare the $n$ photonic qubits in the desired cluster state
 \begin{equation}
 \frac{1}{2^{\frac{n}{2}}}\otimes_{i=1}^{n}(|0\rangle_{i}+\sigma_{i+1}^{z}|1\rangle_{i}),
 \end{equation}
 where $\sigma_{n+1}^{z}\equiv1$.

 In Ref.\,\cite{reve23}, Nielsen showed that 1D cluster states are not universal resource states because any one-way computation performed on 1D cluster states can efficiently be simulated on a classical computer. On the contrary, 2D cluster states have been proven to be universal resource states for one-way quantum computation in the sense that any quantum circuit can be implemented by performing a suitable sequence of single-qubit measurements on a sufficiently large 2D cluster state \cite{reve24}. Therefore, creating large cluster states of dimensions higher than one is essential for implementing the universal quantum computation.

 For the $n\times n$ resonator square lattice, the 2D photonic cluster state can be generated in four steps. First, we entangle the $n$ qubits in each row into a 1D cluster state. Considering that this operation can be performed in parallel for different rows, it can be completed in two steps similar to the 1D case. After this operation, the state of all the photonic qubits is
 \begin{equation}
 \frac{1}{2^{\frac{n^{2}}{2}}}\otimes_{i=1}^{n}[\otimes_{j=1}^{n}(|0\rangle_{i,j}+\sigma_{i,j+1}^{z}|1\rangle_{i,j})],
 \end{equation}
 where $\sigma_{i,j+1}^{z}$ is the Pauli-Z operator for qubit $(i,j+1)$. Second, do the same for the $n$ columns, then we can prepare the $n^{2}$ photonic qubits in the 2D cluster state
 \begin{equation}
 \frac{1}{2^{\frac{n^{2}}{2}}}\otimes_{i,j=1}^{n}(|0\rangle_{i,j}+\sigma_{i,j+1}^{z}\sigma_{i+1,j}^{z}|1\rangle_{i,j}),
 \end{equation}
 where $\sigma_{i,n+1}^{z}\equiv\sigma_{n+1,j}^{z}\equiv1$. Our procedure can be extended to the general case. For a general $d$-dimensional ($d$D) resonator cubic lattice, the $d$D cluster states can be generated in $2d$ steps.

 In one-way quantum computation based on cluster states, calculations are carried out by a series of local measurements in the basis $B(\gamma)=\{|+\gamma\rangle,\; |-\gamma\rangle\}$, where $|\pm\gamma\rangle=(|0\rangle\pm e^{i\gamma}|1\rangle)/\sqrt{2}$ ($\gamma$ is a real number). It is easy to verify that, measuring a qubit in the basis $B(\gamma)$ is equivalent to performing an unitary rotation $U_{\gamma}$ on the qubit followed by a measurement in the basis $\{|0\rangle,\; |1\rangle\}$, where $U_{\gamma}|+\gamma\rangle=|0\rangle$ and $U_{\gamma}|-\gamma\rangle=|1\rangle$. Now, using the experimental architecture shown in Fig.\,3, we give a procedure for measuring the photonic qubit $i$ in the basis $B(\gamma)$. \emph{Step 1}: we let $Q_{i}$ be resonant with $R_{i}$ for a time duration $t=\frac{\pi}{2g_{i}}$, then the system state evolution is
 \begin{equation}
 (\alpha|0\rangle_{i}+\beta|1\rangle_{i})|g\rangle_{i}\rightarrow(\alpha|g\rangle_{i}-i\beta|e\rangle_{i})|0\rangle_{i},
 \end{equation}
 where $\alpha|0\rangle_{i}+\beta|1\rangle_{i}$ is the initial state of photonic qubit $i$. Single-qubit rotations of $Q_{i}$ can be implemented by pulses of microwave applied to the gate driving line. Following the results of Ref.\,[18], different drive frequencies can be chosen to realize rotations around arbitrary axes in the x-z plane, and the rotation angle can be changed easily via varying the microwave pulse length. Let us define $R_{\nu}(\theta)$ ($\nu=x,y,z$) as the rotation of a qubit by an angle $\theta$ around the $\nu$-axis. \emph{Step 2}: we rotate the charge qubit $Q_{i}$ by an angle $\frac{\pi}{2}$ around the z-axis, then the state of $Q_{i}$ evolves into
 \begin{equation}
 R_{z}(\frac{\pi}{2})(\alpha|g\rangle_{i}-i\beta|e\rangle_{i})=\alpha|g\rangle_{i}+\beta|e\rangle_{i}.
 \end{equation}
 After steps 1 and 2, the state of photonic qubit $i$ is perfectly transferred to $Q_{i}$. \emph{Step 3}: we first rotate the charge qubit $Q_{i}$ by an angle $\frac{\pi}{2}-\gamma$ around the z-axis then an angle $\frac{\pi}{2}$ around the x-axis, in this way a single-qubit operation $U_{\gamma}^{Q}=R_{x}(\frac{\pi}{2})R_{z}(\frac{\pi}{2}-\gamma)$ is performed on $Q_{i}$ which satisfies
 \begin{equation}
 \left\{
 \begin{array}{ccc}
 U_{\gamma}^{Q}[\frac{1}{\sqrt{2}}(|g\rangle_{i}+e^{i\gamma}|e\rangle_{i})]&=&|g\rangle_{i},\\
 U_{\gamma}^{Q}[\frac{1}{\sqrt{2}}(|g\rangle_{i}-e^{i\gamma}|e\rangle_{i})]&=&|e\rangle_{i}.\\
 \end{array} \right.
 \end{equation}
 \emph{Step 4}: tune the qubit transition frequency $\epsilon_{i}$ such that $Q_{i}$ is decoupled from $R_{i}$'s half-wave mode but dispersively coupled to $R_{i}$'s full-wave mode. Then the state of $Q_{i}$ can be measured in the basis $\{|g\rangle_{i},\;|e\rangle_{i}\}$ by applying a microwave field $U_{i}^{m}$ to the measurement line. With the help of $Q_{i}$, now we have completed the local measurement of photonic qubit $i$ in the basis $B(\gamma)$ effectively.

 \section{experimental feasibility}

 In this section, we analyze the feasibility of our procedure by some rough calculations based on the practical experimental parameters. First, we give several necessary requirements our architecture parameters must meet. Note that the photon hopping between neighboring resonators will possibly make the relevant photonic qubits run outside of the $\{|0\rangle,\;|1\rangle\}$ manifold. To suppress this effect, the frequency difference $|w-w^{'}|$ of the neighboring resonators must be much larger than the effective photon hopping rate $\kappa_{hop}$ induced by the mediator charge qubit, i.\,e.
 \begin{equation}
 |w-w^{'}|\gg \kappa_{hop}.
 \end{equation}
 The decoherences of the resonators and the charge qubits play significant roles in our procedure. The charge qubits situated in the resonators are used for state initialization and local measurements of the photonic qubits. Because these processes only involve fast single-qubit operations, the decoherences of these inner charge qubits have little effects on our procedure. Each mediator charge qubit participates in the one-way quantum computation process when ``fusing'' its neighboring resonator qubits using a conditional phase gate. To guarantee the high fidelity of the ``fusing'' process, the coherence time of the charge qubit $\tau_{cha}$ must be much longer than the required time of the conditional phase gate $t_{cp}$, i.\;e.
 \begin{equation}
 \tau_{cha}\gg t_{cp}.
 \end{equation}
 Finally, the resonators bear the quantum information for almost the total computation process, thus it is required that the photonic qubits have a coherence time $\tau_{pho}$ much longer than the required time of the total procedure $t_{tot}$, i.\,e.
 \begin{equation}
 \tau_{pho}\gg t_{tot}.
 \end{equation}

 Now, we give a brief evaluation of the involving timescales. The qubit-resonator coupling strength up to $\frac{g}{2\pi}=200\;MHz$  has been realized experimentally in Ref.\,\cite{reve25}. In this case, the required time of a single-qubit rotation can be estimated as $t_{sin}\approx\frac{\pi}{g}=2.5\;ns$. Considering neighboring resonators $R_{i}$ with $\frac{w}{2\pi}=6.6\;GHz$, $R_{i+1}$ with $\frac{w^{'}}{2\pi}=7\;GHz$, and the mediator charge qubit $Q_{i}^{i+1}$ with $\frac{\epsilon_{i}^{i+1}}{2\pi}=8.6\;GHz$, the effective photon hopping rate is approximately $\kappa_{hop}\approx\frac{g^{2}}{|\epsilon_{i}^{i+1}-w|}=2\pi\times20\;MHz$, which is much smaller than the frequency difference $|w-w^{'}|=2\pi\times400\;MHz$. When we implement the conditional phase gate between $R_{i}$ and $R_{i+1}$, to sufficiently suppress the errors induced by the off-resonant transitions, the coupling strength of the Stark-shifted Rabi pulse $\Omega$ should satisfy $|\Omega|\ll min(\frac{2g^{2}}{|\epsilon_{i}^{i+1}-w|},\;\frac{2g^{2}}{|\epsilon_{i}^{i+1}-w^{'}|})=2\pi\times40\;MHz$. Here, we choose $\Omega=2\pi\times4\;MHz$, then the required time of the conditional phase gate is $t_{cp}=\frac{\pi}{\Omega}=125\;ns$. The low-Q full-wave resonator mode with photon decay rate $\kappa_{pho}^{low}=2\pi\times20\;MHz$ can be realized in the experiment by choosing big coupling capacitances for the measurement line \cite{reve20}. Then the local projective measurement of $Q_{i}$ in the basis $\{|g\rangle_{i},\;|e\rangle_{i}\}$ can be implemented in a timescale of $t_{mea}\approx\frac{1}{\kappa_{pho}^{low}}=8\;ns$. To implement one-way quantum computation based on a $d$D resonator cubic lattice consisting of $N$ photonic qubits, we need the time of $2$ single-qubit rotations for the state initialization, the time of $2d$ conditional phase gates for creating the cluster state, and the time of $4$ single-qubit rotations followed by a local measurement for measuring each photonic qubit in a general basis, so the total time of the computation process is approximately
 \begin{equation}
 t_{tot}=2dt_{cp}+N(4t_{sin}+t_{mea})+2t_{sin}=250d+18N+5\;(ns).
 \end{equation}
 On the other hand, the charge qubit with coherence time $\tau_{cha}=1\;\mu s$ and the high-Q resonator mode with coherence time $\tau_{pho}=5\;\mu s$ are reasonable for the practical experimental setup \cite{reve20}. Therefore, considering the necessary conditions (Eqs.\,15-17) and the practical experimental parameters, it is possible to perform one-way quantum computation based on a low dimensional resonator lattice consisting of tens of resonator qubits.

 Our procedure is very robust to the device parameter variations, which are unavoidable in solid-state systems. Although we have assume the same resonance frequency for some resonators and same qubit-resonator coupling strength in the preceding parts, these parameter homogeneities are not necessary in our procedure. Therefore, in the sample fabrication process, the requirements for homogeneity and reproducibility can be relaxed and met with current production technology.

 In conclusion, we propose to construct a scalable one-way quantum computer using on-chip resonator qubits. The unique feature of our architecture is the combination of good scalability and long-lived qubits. With the recent progress in the multi-resonator experiments \cite{reve26}, our proposal may serve as a guide to construct a small quantum computer consisting of more than a handful of qubits.

 This work was supported by the Foundation for the Author of National Excellent Doctoral Dissertation of China (Grant No.\,200524), the Program for New Century Excellent Talents of China (Grant No.\,06-0920), and the National Natural Science Foundation of China (Grant Nos.\,11074307 and 11104353).


\begin{thebibliography}{20}
   \bibitem[1]{reve1} M. A. Nielsen and I. L. Chuang, \emph{Quantum Computation and Quantum Information} (Cambridge University Press, Cambridge, 2000).
   \bibitem[2]{reve2} R. Raussendorf and H. J. Briegel, Phys. Rev. Lett. \textbf{86}, 5188 (2001);
                      H. J. Briegel and R. Raussendorf, Phys. Rev. Lett. \textbf{86}, 910 (2001).
   \bibitem[3]{reve3} P. Walther, K. J. Resch, T. Rudolph, E. Schenck, H. Weinfurter, V. Vedral, M. Aspelmeyer, and A. Zeilinger, Nature (London)
                      \textbf{439}, 169 (2005).
   \bibitem[4]{reve4} R. Prevedel, P. Walther, F. Tiefenbacher, P. B\"{o}hi, R. Kaltenbaek, T. Jennewein, and A. Zeilinger, Nature (London) \textbf{445},
                      65 (2007).
   \bibitem[5]{reve5} M. S. Tame, R. Prevedel, M. Paternostro, P. B\"{o}hi, M. S. Kim, and A. Zeilinger, Phys. Rev. Lett. \textbf{98}, 140501 (2007).
   \bibitem[6]{reve6} M. Borhani and D. Loss, Phys. Rev. A \textbf{71}, 034308 (2005).
   \bibitem[7]{reve7} A. Kolli, B. W. Lovett, S. C. Benjamin, and T. M. Stace, Phys. Rev. Lett. \textbf{97}, 250504 (2006).
   \bibitem[8]{reve8} T. Tanamoto, Y. X. Liu, S. Fujita, X. Hu, and F. Nori, Phys. Rev. Lett. \textbf{97}, 230501 (2006).
   \bibitem[9]{reve9} J. Q, You, X. B. Wang, T. Tanamoto, and F. Nori, Phys. Rev. A \textbf{75}, 052319 (2007).
   \bibitem[10]{reve10} T. Tanamoto, Y. X. Liu, X. Hu, and F. Nori, Phys. Rev. Lett. \textbf{102}, 100501 (2009).
   \bibitem[11]{reve11} M. Hofheinz, H. Wang, M. Ansmann, R. C. Bialczak, E. Lucero, M. Neeley, A. D. O'Connell, D. Sank, J. Wenner, J. M. Martinis, and
                        A. N. Cleland, Nature (London) \textbf{459}, 546 (2009).
   \bibitem[12]{reve12} H. Wang, M. Mariantoni, R. C. Bialczak, M. Lenander, E. Lucero, M. Neeley, A. D. O'Connell, D. Sank, M. Weides, J. Wenner, T. Yamamoto,
                        Y. Yin, J. Zhao, J. M. Martinis, and A. N. Cleland, Phys. Rev. Lett. \textbf{106}, 060401 (2011).
   \bibitem[13]{reve13} A. Galiautdinov, A. N. Korotkov, and J. M. Martinis, e-print arXiv: 1105. 3997.
   \bibitem[14]{reve14} A. Galiautdinov, e-print arXiv: 1103. 4641.
   \bibitem[15]{reve15} F. W. Strauch, e-print arXiv: 1108. 2984.
   \bibitem[16]{reve16} D. I. Tsomokos, S. Ashhab, and F. Nori, Phys. Rev. A \textbf{82}, 052311 (2010).
   \bibitem[17]{reve17} R. Harris \emph{et al.}, Phys. Rev. B \textbf{82}, 024511 (2010).
   \bibitem[18]{reve18} A. Blais, R. S. Huang, A. Wallraff, S. Girvin, and R. J. Schoelkopf, Phys. Rev. A \text{69}, 062320 (2004).
   \bibitem[19]{reve19} D. I. Schuster, A. A. Houck, J. A. Schreier, A. Wallraff, J. M. Gambetta, A. Blais, L. Frunzio, J. Majer, B. Johnson, M. H. Devoret,
                        S. M. Girvin, and R. J. Schoelkopf, Nature (London) \textbf{445}, 515 (2007).
   \bibitem[20]{reve20} B. R. Johnson, M. D. Reed, A. A. Houck, D. I. Schuster, Lev S. Bishop, E. Ginossar, J. M. Gambetta, L. DiCarlo, L. Frunzio, S. M. Girvin,
                        and R. J. Schoelkopf, Nat. Phys. \textbf{6}, 663 (2010).
   \bibitem[21]{reve21} F. W. Strauch, K. Jacobs, and R. W. Simmonds, Phys. Rev. Lett. \text{105}, 050501 (2010).
   \bibitem[22]{reve22} P. J. Leek, M. Baur, J. M. Fink, R. Bianchetti, L. Steffen, S. Filipp, and A. Wallraff, Phys. Rev. Lett. \textbf{104}, 100504 (2010).
   \bibitem[23]{reve23} M. A. Nielsen, Reports on Mathematical Physics \textbf{57}, 147 (2006).
   \bibitem[24]{reve24} M. Van den Nest, A. Miyake, W. D\"{u}r, and H. J. Briegel, Phys. Rev. Lett. \textbf{97}, 150504 (2006).
   \bibitem[25]{reve25} L. Dicarlo, M. D. Reed, L. Sun, B. R. Johnson, J. M. Chow, J. M. Gambetta, L. Frunzio, S. M. Girvin, M. H. Devoret, and R. J. Schoelkopf,
                        Nature (London) \textbf{467}, 574 (2010).
   \bibitem[26]{reve26} Matteo Mariantoni, H. Wang, Radoslaw C. Bialczak, M. Lenander, Erik Lucero, M. Neeley, A. D. O'Connell, D. Sank, M. Weides, J. Wenner, T.
                        Yamamoto, Y. Yin, J. Zhao, John M. Martinis, and A. N. Cleland, Nat. Phys. \textbf{7}, 287 (2011); H. Wang, Matteo Mariantoni, Radoslaw C. Bialczak, M. Lenander, Erik Lucero, M. Neeley, A. D. O'Connell, D. Sank, M. Weides, J. Wenner, T. Yamamoto, Y. Yin, J. Zhao, John M. Martinis, and A. N. Cleland, Phys. Rev. Lett. \textbf{106}, 060401 (2011).
\end{thebibliography}
\end{document}